\documentclass{iau}
\usepackage{float}
\usepackage{graphicx}
\title[Seven Pulsars in Binary System] %% short title %%
{Seven Pulsars in Binary Systems above the Spin-up Line} %% full title %%

\author[Y.Y. Pan \& N. Wang]  %% short author list %%
{Y.Y. Pan
 \and N. Wang}

\affiliation{Xinjiang  Astronomical Observatories, Chinese Academy of
Sciences, Xinjiang 830011, China\\ Email: {\tt panyuanyue@xao.ac.cn}}
\pubyear{2012}
\volume{291}  %% insert here IAU Symposium No.
\jname{\mbox{Neutron Stars and Pulsars: Challenges and Opportunities after 80 years}}
\editors{J.~van Leeuwen, ed.}
\begin{document}

\maketitle

%% -- Abstract ----------------------------------
\begin{abstract}
Using data from the ATNF pulsar catalogue, 186 binary pulsars are shown
in the magnetic
field versus spin period (B-P) diagram, and their relationship to the
spin-up line is
investigated. Generally speaking, pulsars in binary systems should be
below the spin-up
line when they get enough accretion mass from their companions. It is
found that there are
seven binary pulsars above the spin-up line.  Based on the
parameters of these seven
binary systems, we describe possible reasons why they are above the
spin-up line.

\end{abstract}

% add below any authors, subjects and objects for indexing
%   add more lines if necessary
%   but leave all lines commented out
%\index[author]{LastName1, Initials|textbf}
%\index[author]{LastName2, Initials|textbf}
%\index[subject]{Keyword1}
%\index[subject]{Keyword2}
%\index[object]{Object1}
%\index[object]{Object2}

\firstsection % if your document starts with a section,
\section{Introduction}
A binary pulsar system is a pulsar with a companion, often a white dwarf,
neutron star or massive star. When a neutron star is formed from the supernova,
it has a high magnetic field of around $10^{11-13}$\,G and slow spin
period of around 0.1$-$10\,s.
%Utilizing binary evolution calculations, it has been studied a number of systems,
%finding that magnetic field strength decrease with the amount of accreted matter.
In a binary system, with the accretion mass of $0.1\sim 0.2 M_{\odot}$ from the
companion, a neutron star will be spun up to several milliseconds, while its
magnetic field will decrease to $\sim$10$^{8-9}$\,G
(\cite[Bhattacharya \& van den Heuvel 1991]{bha91};\cite[Manchester 2004]{man04};
\cite[Stairs 2004]{sta04}; \cite[Manchester, Hobbs, Teoh \& Hobbs 2005]{man05};
\cite[Wang, Zhang \& Zhao et al. 2011]{wang11}; \cite[Zhang, Wang \& Zhao et al. 2011]{zhang11}).
During the accretion, the flow drag the field lines asides to dilute the polar field
strength (\cite[Zhang \& Kojima 2006]{zhang06}).
When a neutron star gets to its minimum spin period to which such a spin up
proceeds in an Eddington-limited accretion, we can get the spin-up line
(\cite[Bhattacharya \& van den Heuvel 1991]{bha91}):
%\begin{equation}
$B_9^{\frac{6}{7}}=\frac{P}{2.2}
R_6^{-\frac{18}{7}}m^{\frac{8}{7}}\times10^3$,
%\end{equation}
where $B_9$ is the magnetic field in units of $10^9$\,G, $R_6$ is the stellar radius
in units of $10^6$cm, $m$ is in unit of  solar mass. After the accretion phase
finishing, the radio emission of the fast rotating neutron star can be detected
as millisecond pulsar whose spin period is less than 20 milliseconds
(\cite[Alpar, Cheng, Ruderman \& Shaham 1982]{alp82}; \cite[Tauris 2012]{tau12}).
%

%In binary system, the average
%mass for millisecond pulsars and others are $M=1.57\pm0.35M_\odot$ and $M=1.37\pm0.23M_\odot$, respectively\cite{zhang11}.
%In order to get the spin-up lien in the figure, we choose $R_6=1$ and
%$m=1.4M_{\odot}$. All binary pulsars should be under this line if
%sufficient mass is accreted.

\section{Evolution of Pulsar in Binary system}
\begin{figure}[tb]
\centering
\includegraphics[width=3.2in]{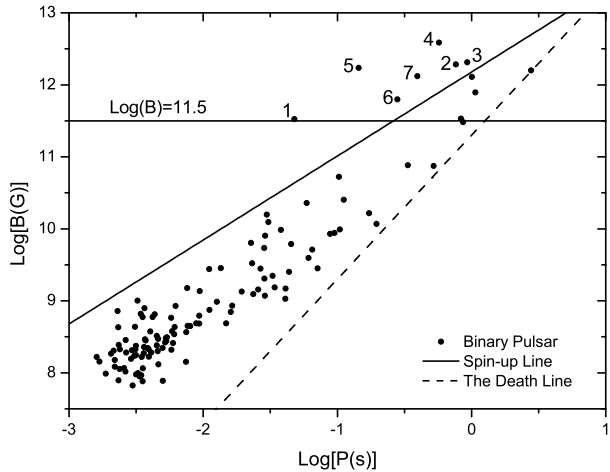}
\caption{Magnetic field versus spin period diagram for 186 pulsars in binary
system, data from ATNF pulsar catalogue. Seven pulsars are above the spin-up
line which are signed by arranged number. The horizontal line is the magnetic
field $B=10^{11.5}$\,G. The solid and dash lines are the spin-up line and death
line.}
%\label{fig1}
%\end{center}
\end{figure}
Up to now 186 pulsars (including 136 millisecond ones) have been found
in binary
systems (data from ATNF pulsar catalogue). Fig.\ 1 shows us their distribution in
the B-P diagram. In binary systems, pulsars will evolve below the spin-up line
after accreting a sufficient amount of mass from their companions. However, there
are seven binary pulsars which lie above the spin-up line, as shown in
Fig.\ 1. We arranged
them into two groups according to their companion masses: the first one with the
massive companions ($M>4.0M_{\odot}$, No.\ $1\sim4$) and the second one with the
degenerate stars (white dwarf or neutron star, No.\ $5\sim7$). The parameters of
seven binary pulsars are listed in Table 1.

\begin{table}[b]
  \begin{center}
  \caption{Information of 7 binary pulsars above the spin-up line}
  \label{tab1}
 {\begin{tabular}{lccccccccccccccl}\hline
   No.& Name& $P/ms$ &$P_{orb}/d$   &e    &$M_C/M_{\odot}$        &$\tau/yr$ &$B/G $ &type\\
\hline
1& B1259-63& 47.763&         1236.724&   0.8699&     4.14&       332000& 3.34E11&NSMS\\
   2&J1638-4725&    763.933&    1940.9&     0.955&      8.08&       2.53E6& 1.93E12&NSMS\\
   3&J0045-7319&    926.276&    51.1695&    0.8079&     5.27&       3.29E6& 2.06E12&NSMS\\
   4&J1740-3052&    570.31&     231.0297&   0.5789&     15.82&      354000& 3.86E12&NSMS\\
   5&J1906+0746&    144.072&    0.166&      0.0853&     1.37&       113000& 1.73E12&DNS\\
   6&B1820-11&  279.829&        357.762&    0.7946&     0.78&       3.22E6& 6.29E11&NSWD\\
   7&J1141-6545&    393.899&    0.1977&     0.1719&     1.02&       1.45E6& 1.32E12&NSWD\\
    \hline
  \end{tabular}
  }
 \end{center}
\vspace{1mm}
\end{table}

The common characteristics of the first group (No.\  1 to 4) are their
massive and non-recycled companions. They all have high eccentricities
and long orbital
periods. The accretion phase has not yet started. Due to these
characteristics, it can be said that they are un-recycled pulsars whose companions are still
in the main sequence in binary systems. Or they are on the way of evolution, where the first
born pulsars are experiencing the spin-down with no accretion.

The second group (No.\ 5 to 7) includes the recycled pulsars. PSR~J1906+0746
(No.\ \ 5 in the Fig.\ 1)
with mass ($1.25M_\odot$) has a neutron star companion of mass $1.37M_\odot$. Comparing
the mass of the two neutron stars, it is inferred that the heavier one is a
recycled pulsar and the lighter one is a non-recycled pulsar. From the evolution history
of double neutron stars, the heavier progenitor star explodes first to form a neutron star,
and then the lighter one evolves until to its supernova explosion. During the evolution of the
second star, the first formed neutron star will accrete matter, leading to its recycling. The short
characteristic age  of J1906+0746 ($\tau$=113000\,yr) indicates that it is a recently formed
young pulsar after the core collapse, which is the reason why its B-P position lies above the
spin-up line.

B1820-11 (No.\ 6 in Fig.\ 1) possesses a slightly  massive white dwarf ($M_c=0.78 M_{\odot}$)
as its companion. From its parameters ($\tau$=3.22\,Myr, B=6.29$\times
10^{11}$\,G and
$P$=279.829\,ms), we suggest that it is a young recycled pulsar.
%And its companion is
%a recycled star as a white dwarf. %The formation can be understood like this: in the binary
%system, the heavier star explode to be a helium core. During the evolution of the second
%star, the helium core accretes the matter through the envelope. Then the helium collapses
%to a white dwarf and the second
%
If we reconsider its radius as 15\,km or 20\,km instead of 10\,km in the spin-up line equation,
then its magnetic field is about $3.0\times 10^{11}$\,G and $1.6 \times 10^{11}$\,G, respectively.
Therefore, in Fig.\ 1, the position of B1820-11 with the new magnetic field value will lie
below the spin-up line.
The evolution history of this pulsar can be understood in this way: the initial magnetic
field of neutron star can be as high as $B \sim 10^{13}$\,G, and it can evolve from the
position of long spin period to that of short spin period after the neutron star accretes
about $\sim0.001 M_\odot$, while one to two magnitude orders of magnetic field has been deducted.

J1141-6545 (No.\ 7 in Fig.\ 1) is a pulsar of mass $1.3M_\odot$ with an $1.02M_\odot$ optical
white dwarf as its companion in binary system. With the short orbital period
($P_{orbit}=0.1977 d$) and low eccentricity ($e=0.1719$), it can be derived that this
pulsar acquired the accretion mass easily and followed up the recycled process. There
is about $0.001 - 0.01 M_\odot$ accretion mass added to this pulsar that can lead to
its magnetic field deduce two magnitude orders from its initial values. Following the
similar procedure of binary pulsar B1820-11 (No.\ 6), by setting the neutron star radius
as large as R=20 km, the magnetic field of J1141-6545  will be about $10^{11.6}$\,G, which
makes this source just below the spin-up line as a new born recycled pulsar.
Therefore, the evolution picture of this pulsar can be depicted like this: the progenitor
of J1141-6545 may be a star with the strong magnetic field $\sim 10^{13.6}$\,G, and its
field decays two magnitude orders with the accreting mass of  about $0.001M_\odot$.
%
%The other possibility for this source is that the pulsar may experience the accreting
%induced collapse of a white dwarf, so the evolution history of J1141-6545 does not
%follow the conventional route of the recycled PSR.

\section{Conclusion}
In a binary system, with enough accreting matter from the companions, a pulsar
should be below the spin-up line. With the distribution of 186 binary pulsars
in B-P diagram, it is noticed that seven pulsars are above the spin-up line. Four of
them have massive companions ($M>4.0M_\odot$) and are young pulsars which are
quickly spinning down.  They have not started their recycling processes.
The other three binary pulsars with recycled companions have not experienced the recycled
processes: one system is a double neutron system. The observed pulsar is a young
one with a recycled neutron star. The other two systems include degenerate stars (NS+WD),
where the pulsars can be understood as newly formed recycling
pulsars at the Eddington rate; and their B-P positions can be
shifted to just below the spin-up line by assuming a different
 neutron star radius, e.g.\ $R=20$\,km.

~\\
This work is supported by National
Basic Research Program of China (973 Program 2009CB824800), China Ministry of Science and
Technology under State Key Development Program for Basic Research
(2012CB821800), NSFC10773017,  NSFC11173034, and Knowledge Innovation Program of CAS KJCX2-
YW-T09.

%NBRPC2009CB824800,
%NBRPC2012CB821800,  NSFC10773017,  NSFC11173034, and CKIP(KJCX2-YW-T09).
%
%
%Chinese National Science Foundation
%through grant No. 10773005, National Basic Research Program of
%China (973 Program 2009CB824800), China Ministry of Science and
%Technology under State Key Development Program for Basic Research
%(2012CB821800), Knowledge Innovation Program of CAS KJCX2-
%YW -T09, Xinjiang Natural Science Foundation No. 2009211B35, the
%Key Directional Project of CAS and NSFC under projects 10173020,
%10673021, 10773005, 10778631 and 10903019.

\end{document}